\documentclass[english,aps, manuscript, twocolumn, showpacs, 10pt]{revtex4}

\usepackage[T1]{fontenc}
\usepackage[latin9]{inputenc}
\usepackage{color}
\usepackage{babel}

\usepackage{bm}
\usepackage{amsmath}
\usepackage{graphicx}
\usepackage{amssymb}
\usepackage{esint}
\usepackage[unicode=true, pdfusetitle,
 bookmarks=true,bookmarksnumbered=false,bookmarksopen=false,
 breaklinks=false,pdfborder={0 0 0},backref=false,colorlinks=true]
 {hyperref}

\makeatletter

\@ifundefined{textcolor}{}
{%
 \definecolor{BLACK}{gray}{0}
 \definecolor{WHITE}{gray}{1}
 \definecolor{RED}{rgb}{1,0,0}
 \definecolor{GREEN}{rgb}{0,1,0}
 \definecolor{BLUE}{rgb}{0,0,1}
 \definecolor{CYAN}{cmyk}{1,0,0,0}
 \definecolor{MAGENTA}{cmyk}{0,1,0,0}
 \definecolor{YELLOW}{cmyk}{0,0,1,0}
 }

\def\bra#1{{\langle#1|}}\def\ket#1{{|#1\rangle}}\def\bracket#1#2{{\langle#1|#2\rangle}}

\@ifundefined{showcaptionsetup}{}{\PassOptionsToPackage{caption=false}{subfig}}\usepackage{subfig}\makeatother

\usepackage{babel}\addto\shorthandsspanish{\spanishdeactivate{~<>}}

\usepackage{bbm}

\makeatother

\begin{document}

\title{The classical skeleton of open quantum chaotic maps}

\author{Lisandro A. Raviola$^{1,2}$, Alejandro
M.F. Rivas$^{2}$, and Gabriel G. Carlo$^{2}$}

\email{raviola@tandar.cnea.gov.ar}

\affiliation{$^{1}$Instituto S\'abato (UNSAM - CNEA), Av. Gral. Paz 1499, B1650KNA
San Mart\'\i n, Buenos Aires, Argentina}

\affiliation{$^{2}$Departamento de F\'\i sica, CNEA, Av. del Libertador 8250, C1429BNP
Buenos Aires, Argentina }

\date{\today}


\begin{abstract}
We have studied two complementary decoherence measures purity
and fidelity for a generic diffusive noise in two different chaotic systems (the 
baker and the cat maps). For both quantities, we have found classical structures in 
quantum mechanics -- the scar functions -- that are specially stable when 
subjected to environmental perturbations. We show that these quantum states constructed 
on classical invariants are the most robust significant quantum distributions 
in generic dissipative maps. 
\end{abstract}

\maketitle

\section{Introduction}

In quantum chaos (i.e. the study of quantum systems whose classical 
counterparts are chaotic), the complete description of the eigenfunctions 
is still an open problem. While there is a semiclassical method for 
obtaining the eigenstates of an integrable system (the well-known EBK/WKB 
quantization scheme \cite{Brack,Maia}), the same
problem in the chaotic case has proven harder to solve. One of the major 
advances in this sense has been Gutzwiller's theory of periodic orbits
for the quantum chaotic spectra \cite{Gutzwiller}, devised in the 
early $70$'s. 

A recently developed approach allows to obtain all the information 
of a generic quantum system by just using the shortest periodic orbits 
of its classical counterpart \cite{ScarFunctions}. This formalism has been successfully
applied to the description of the eigenstates of many chaotic quantum
systems like for example, the Bunimovich's billiard \cite{ScarsBilliard}, the cat
map \cite{ScarsCat} and the baker map \cite{ScarsBaker}. These results
suggest that states built using only this classical information (the so-called \emph{scar
functions}) constitute the skeleton of the eigenstates of any quantum
chaotic system.

On the other hand, taking into account the restoration of the classical
dynamics induced by decoherence in an open quantum system \cite{DecoherenceChaos} 
and the behavior of distributions against external perturbations \cite{Jacquod,Jalabert,Bonanza} 
we can ask ourselves: are scar functions, as a consequence of their classical 
content, more robust than other states facing an environment-induced
decoherence? Trying to answer this question, in \cite{Raviola} we have studied 
a very simple model of a chaotic quantum system interacting with an
environment that produces dissipation and decoherence (an \emph{open quantum map}). 
In this article, we exhaustively expand the results presented there by studying different 
open maps. This leads us to our main result, i.e. we show here that scar functions 
are the most robust significant quantum distributions 
in generic dissipative maps. 

In this paper we study the behavior of classically motivated states 
(by analyzing two different measures, namely the purity and fidelity) 
corresponding to two paradigmatic systems in quantum chaos,
the baker map and the cat map on the torus \cite{QuantumMaps,Arnold}. We 
introduce decoherence by means of a diffusive noise model \cite{GarciaMata}.
We have organized this work as follows. In section 2, we describe the main theoretical 
tools we need throughout our investigations (open quantum maps and scar functions). 
In Section 3 we define the maps considered. Section 4 presents our results. 
In section 5 we come to some conclusions.

\section{Theoretical tools}

Open quantum maps are the simplest systems that capture all the essential features of 
chaotic dynamics and dissipation. As such, they are the ideal testbed for studying 
the effect of the environment on quantum distributions. Our main 
interest is focused on determining how the classical information emerges from the 
quantum structures. For that purpose we will make use use of the scar functions, whose 
construction we explain at the end of this Section.                   

\subsection{Open quantum maps}

The quantization of maps on a compact phase space proceeds in two
stages: a kinematic one, which establishes the Hilbert space appropriate
to the phase space geometry, and a dynamical one, which consists in
defining a suitable quantum operator corresponding to the classical
dynamics. Finally, Kraus operators take into account the effect of the 
environment.

\subsubsection{Kinematics}

In this work, the classical phase space associated with the systems
under investigation is the 2-dimensional torus $\mathbb{T}^{2}=\mathbb{R}^{2}/\mathbb{Z}^{2}$,
consisting of a square of unit side with opposite sides identified.
Points in this space have coordinates $(q,p)\in[0,1)\times[0,1)$.
To be compatible with the phase space geometry, quantum wave functions
must be periodic in both position and momentum (up to a phase)\begin{eqnarray}
\psi(q+1) & = & \exp\left(-\mathrm{i}\,2\pi\chi_{q}\right)\ \psi(q)\label{eq:periodic_cond}\\
\tilde{\psi}(p+1) & = & \exp\left(\mathrm{i}\,2\pi\chi_{p}\right)\ \tilde{\psi}(p)\end{eqnarray}
 with \begin{equation}
\tilde{\psi}(p)=\frac{1}{\sqrt{2\pi\hbar}}\int_{-\infty}^{\infty}dq\ e^{-\frac{i}{\hbar}qp}\ \psi(q).\label{eq:fourier_transf}\end{equation}
 The phases $2\pi\chi_{q},2\pi\chi_{p}$ are called \emph{Floquet's
angles}, with $0\leq\chi_{q},\chi_{p}<1$. Periodicity in $q$ and
$p$ implies \begin{eqnarray}
2\pi\hbar N & = & 1\label{eq:torus_hilbert_dim}\end{eqnarray}
This means that the Hilbert space of wave functions $\mathcal{H}_{N}$,
the quantum counterpart of the classical compact phase space, is effectively
finite-dimensional with dimension $N$. In this context, the semi-classical
limit $\hbar\rightarrow0$ is equivalent to taking $N\rightarrow\infty$.$ $

The usual canonical commutation relations between position and momentum
operators (which are the generators of \emph{infinitesimal} translations
in phase space) do not hold in the finite-dimensional case, so we
can't define position and momentum operators $\hat{q},\hat{p}$. However,
we can define \emph{finite }displacement operators $\hat{U}$ and
$\hat{V}$, whose form is analogous to that of the infinite-dimensional
case. These operators are unitary and its eigenvectors form a basis
for $\mathcal{H}_{N}$ \cite{Schwinger,Miquel,OzorioRep,TracyPhD}.

The basis of position vectors $\ket{q_{j}}$ for this space will be
defined from the eigenvectors of the momentum displacement operator
$\hat{V}$\begin{equation}
\hat{V}\ket{q_{j}}=\exp\left[\frac{2\pi\mathrm{i}}{N}\left(j+\chi_{q}\right)\right]\ket{q_{j}},\quad j\in[0,N-1]\label{eq:momentum_displacement}\end{equation}
The position in phase space associated with $\ket{q_{j}}$ is \begin{equation}
q_{j}=\frac{j+\chi_{q}}{N}.\label{eq:q_value}\end{equation}
Analogously, the momentum vectors $\ket{p_{k}}$ satisfy 

\begin{eqnarray}
\hat{U}\ket{p_{k}} & = & \exp\left[-\frac{2\pi\mathrm{i}}{N}\left(k+\chi_{p}\right)\right]\ket{p_{k}}\label{eq:position_displacement}\\
p_{k} & = & \frac{k+\chi_{p}}{N},\quad k\in[0,N-1]\label{eq:p_value}\end{eqnarray}
where $\hat{U}$ is the position displacement operator. 

These bases are related by means of a discrete Fourier transform\begin{equation}
\bracket{p_{k}}{q_{j}}=N^{-\frac{1}{2}}\exp\left[-\frac{2\pi\mathrm{i}}{N}\left(j+\chi_{q}\right)\left(k+\chi_{p}\right)\right]\equiv\hat{F}_{N}^{j,k}\label{eq:DFT}\end{equation}
From the previous equations, it can be shown that \cite{OzorioRep,TracyPhD,Schwinger,Miquel}
\begin{eqnarray}
\hat{U}\ket{q_{j}} & = & \ket{q_{j+1}}\\
\hat{U}^{N}\ket{q_{j}} & = & \ket{q_{j+N}}=\exp\left(-\mathrm{i}2\pi\chi_{q}\right)\ket{q_{j}}\\
\hat{V}\ket{p_{k}} & = & \ket{p_{k+1}}\\
\hat{V}^{N}\ket{p_{k}} & = & \ket{p_{k+N}}=\exp\left(\mathrm{i}2\pi\chi_{p}\right)\ket{p_{k}}\end{eqnarray}
These operators satisfy the relation\begin{equation}
\hat{U}^{j}\hat{V}^{k}=\hat{V}^{k}\hat{U}^{j}\exp\left(\frac{2\pi\mathrm{i}}{N}jk\right)\label{eq:conmutationUV}\end{equation}
 With them, we can define a discrete version of the phase space displacement
operator
\begin{equation}
\hat{T}_{j,k}=\frac{1}{\sqrt{N}}\exp\left(\frac{\mathrm{i}\pi}{N}jk\right)\hat{V}^{j}\hat{U}^{k}
\label{eq:translation_op}
\end{equation}
with the property $\hat{T}_{j,k}^{\dagger}=\hat{T}_{-j,-k}$. The
set $\left\{ \hat{T}_{j,k}\right\} _{j,k=0}^{N^{2}-1}$ of finite
displacements forms a basis for the Hilbert space $\mathcal{H}_{N^{2}}=\mathcal{H}_{N}\otimes\mathcal{H}_{N}^{*}$
of linear operators on $\mathcal{H}_{N}$ (Liouville space) with the
Hilbert-Schmidt inner product\begin{equation}
(\hat{A},\hat{B})=\mathrm{Tr}\left(\hat{A}^{\dagger}\hat{B}\right)\label{eq:hilbert_schmidt_prod}\end{equation}
This basis of $N^{2}$ displacement operators satisfies\begin{equation}
\mathrm{Tr}\left(\hat{T}_{j,k}^{\dagger}\hat{T}_{j',k'}\right)=\delta_{j,j'}\delta_{k,k'}\label{eq:orthonorm_translation}\end{equation}
 so it constitutes a complete orthonormal set.

\subsubsection{Dynamics}

By virtue of the finite dimension of $\mathcal{H}_{N}$, the quantum
dynamics is given by a unitary $N\times N$ matrix $\hat{U}_{N}$
and the system state evolution is obtained by means of a straightforward
matrix multiplication. In our case, this matrix will be the quantization
of a classical (chaotic) map $P$ on $\mathbb{T}^{2}$. This means
that, given the mapping $P(q,p)$, there exists a sequence of unitary
operators $\hat{U}_{N}$ (called \emph{quantum map}) acting on $\mathcal{H}_{N}$
such that the so-called Egorov property is fulfilled, i.e.\begin{equation}
\lim_{N\rightarrow\infty}\left\Vert \hat{U}_{N}^{-1}\mathrm{Op}(f)\hat{U}_{N}-\mathrm{Op}\left(f\circ P\right)\right\Vert =0\qquad\forall f\in C^{\infty}(\mathbb{T}^{2})\label{eq:egorov_property}\end{equation}
$\mathrm{Op}(f)$ represents the Weyl quantization of the observable
$f$ \cite{OzorioRep}.

\subsubsection{Environment}

If the quantum system interacts with an environment, the elements
(kets) of $\mathcal{H}_{N}$ no longer represent its state. In this
situation, all that can be said about the system at time $t$ is encoded
in its associated density operator $\hat{\rho}_{t}$ \cite{Cohen-Tannoudji,Nielsen,Preskill}.
The evolution of $\hat{\rho}_{t}$ is given, under sufficiently general
conditions, by a completely positive, trace-preserving map $\boldsymbol{S}$
of density matrices into density matrices called \emph{superoperator
}or \emph{quantum operation }\cite{Kraus,Nielsen,Preskill}\emph{.}

In this context, we define an \emph{open quantum map} $\boldsymbol{S}$
\cite{Bianucci,GarciaMata} as a map whose action can be written in
the form of a product of two superoperators \begin{equation}
\hat{\rho}_{t+1}=\boldsymbol{S}(\hat{\rho}_{t})=\boldsymbol{D}_{\varepsilon}\,\boldsymbol{M}(\hat{\rho}_{t})\label{eq:openmap}\end{equation}
$\boldsymbol{M}(\hat{\rho}_{t})=\hat{M}\hat{\rho}_{t}\hat{M}^{\dagger}$
is a map that generates the unitary evolution of the system, $\hat{M}$
being the evolution operator that acts upon elements of the Hilbert
space associated to the non-interacting system. $\boldsymbol{D}_{\varepsilon}$
is a superoperator that models the interaction between system and
environment according to a set of parameters $\varepsilon$ related
to the specific type of interaction. This last superoperator is responsible
for introducing noise in the --otherwise unitary-- system evolution.

In the present work, $\hat{M}$ will be the quantization of a classically
chaotic map acting on the torus. In particular, we will consider two
completely chaotic maps: the baker map and the cat map \cite{QuantumMaps,Arnold}.
Besides, the noise superoperator will be expressed according to the
Kraus representation \cite{Kraus,Nielsen,Preskill}, which in general can 
be expressed as 
\begin{equation}
\boldsymbol{D}_{\varepsilon}(\hat{\rho}_{t})=\sum_{i=0}^{N^{2}-1}\hat{K}_{i}
\hat{\rho}_{t}\hat{K}_{i}^{\dagger}\label{eq:krausrep}
\end{equation}
with $\sum_{i}\hat{K}_{i}^{\dagger}\hat{K}_{i}=\mathbbm{1}$ in order
to preserve the trace of $\hat{\rho}_{t}$. The way the system interacts
with the environment will be completely determined by the Kraus operators
$\hat{K}_{i}$. In the following Section, we will define these operators explicitly.
An example of the action of the environment in phase space can be seen in 
Figs. \ref{fig:figure1} and \ref{fig:figure2}.

\subsection{Scar functions}

The semiclassical theory of short periodic orbits \cite{ScarFunctions} 
is a formalism that allows to
obtain all the quantum information of a chaotic Hamiltonian system
in terms of a very small number of short periodic orbits. The main
elements in this theory are the so-called \emph{scar functions}. These
are wavefunctions highly localized in the neighborhood of the classical
periodic orbits and on their stable and unstable manifolds, satisfying
a Bohr-Sommerfeld quantization condition along the trajectory. They
are defined for Hamiltonian flows as \begin{equation}
\ket{\phi_{scar}}=\int_{-T}^{T}dt\ \cos\left(\frac{\pi t}{2T}\right)e^{\frac{i}{\hbar}\left(E_{BS}-\hat{H}\right)t}\ket{\phi_{tube}}\label{eq:scarshamiltonian}\end{equation}
where $\hat{H}$ is the system's Hamiltonian, $T$ is of the order
of Ehrenfest's time, and $\ket{\phi_{tube}}$ is a wavefunction localized
on the periodic orbit with Bohr-quantized energy $E_{BS}$. In \cite{ScarsCat,ScarsBaker}
the formalism has been adapted to quantum maps on the torus, and the resulting
formula for scar functions is given in terms of a sum\begin{equation}
\ket{\phi_{scar}^{maps}}=\sum_{t=-T}^{T}\cos\left(\frac{\pi t}{2T}\right)e^{\frac{i}{\hbar}E_{BS}t}\hat{\ U^{t}}\ket{\phi_{POM}^{maps}}\label{eq:scarsmap}\end{equation}
where $\hat{U}$ is the evolution operator of the quantum map and
$\ket{\phi_{POM}^{maps}}$ (called \emph{Periodic Orbit Mode }or POM\emph{)
}is a sum of coherent states on the torus centered at the fixed points
of a given periodic orbit, each one having a phase. In this case,
the Ehrenfest time is $T=\frac{\ln N}{\lambda}$, $\lambda$ is the
Lyapunov exponent of the map and $N$ is the Hilbert space dimension. 

As an example, the upper right panels of Figs. \ref{fig:baker_figs}
and \ref{fig:cat_figs} show Husimi representations of scar functions
constructed for the baker map and for the cat map, respectively.
It's clearly visible the enhancement of probability that these wavefunctions
have on the corresponding periodic orbit and its stable and unstable
manifolds.

\section{Systems}

\subsection{Baker map}

The first model of chaotic dynamics we consider is the baker map $B:\ \mathbb{T}^{2}\rightarrow\mathbb{T}^{2}$,
given by the transformation \cite{Arnold,QuantumMaps} \begin{equation}
(q',p')=B(q,p)=[2q-\lfloor2q\rfloor,(p+\lfloor2q\rfloor)/2]\label{eq:bakermap}\end{equation}
where $\lfloor x\rfloor$ stands for the integer part of $x$. This
transformation is an area-preserving, uniformly hyperbolic, piecewise-linear
and invertible map with Lyapunov exponent $\lambda=\ln2$. The vertical
(horizontal) lines $q=q_{0}\ (p=p_{0})$ represent the stable (unstable)
manifolds.

The phase space has a very simple Markov partition consisting of two
regions ($q<1/2$ and $q\geq1/2$) associated with the symbols $0$
and $1$, for which there is a complete symbolic dynamics. The action
of the map upon symbols can be understood by means of the binary expansion
of the coordinates \begin{equation}
(p|q)=\ldots\nu_{-1}\cdot\nu_{0}\nu_{1}\ldots\stackrel{B}{\longrightarrow}(p\prime|q\prime)=\ldots\nu_{-1}\nu_{0}\cdot\nu_{1}\ldots\label{eq:shift}\end{equation}
 where $q=\sum_{i=0}^{\infty}\nu_{i}2^{-(i+1)}$ and $p=\sum_{i=-1}^{-\infty}\nu_{i}2^{i}$.
Then, a periodic orbit of period $L$ can be represented by a binary
string $\bm{\nu}$ of length $L$. The coordinates of the first trajectory
point $(q_{0},p_{0})$ on the periodic orbit can be obtained explicitly
in terms of the binary string as $q_{0}=\cdot\bm{\nu\nu\nu}\ldots=\nu/(2^{L}-1)$
and $p_{0}=\cdot\bm{\nu^{\dagger}\nu^{\dagger}\nu^{\dagger}}\ldots=\nu^{\dagger}/(2^{L}-1)$,
where $\nu$ is the integer value of the string $\bm{\nu}$ which
represents a binary number, and $\bm{\nu^{\dagger}}$ is the string
formed by all $L$ bits of $\bm{\nu}$ in reverse order. The other
trajectory points can be easily calculated by iterations of the map
or by cyclic shifts of $\bm{\nu}$.

The unitary operator $\hat{M}$ that performs the closed quantum evolution
is given in position representation by \cite{Voros,Saraceno} \begin{equation}
\hat{M}=\hat{F}_{N}^{\dagger}\left(\begin{array}{cc}
\hat{F}_{N/2} & \mathbb{O}\\
\mathbb{O} & \hat{F}_{N/2}\end{array}\right)\label{quantumbaker}\end{equation}
where $\hat{F}_{N}$ is the $N$-dimensional Fourier transform operator
whose matrix elements were defined in \eqref{eq:fourier_transf}.
Throughout the paper we assume for the quantum baker map a phase space
with anti-symmetric boundary conditions ($\chi_{q}=\chi_{p}=1/2$)
in order to preserve the classical map symmetries \cite{Saraceno}.

\subsection{Cat map}

Another simple model with strongly chaotic dynamics on $\mathbb{T}^{2}$
is the cat map \cite{Arnold,QuantumMaps}. It's an invertible, area-preserving
canonical transformation $A$ whose matrix has integer entries, and
with $\mathrm{Tr}(A)>2$ to ensure hyperbolicity. A common choice
for $A$ is \begin{equation}
\left(\begin{array}{c}
q'\\
p'\end{array}\right)=A\left(\begin{array}{c}
q\\
p\end{array}\right)=\left(\begin{array}{cc}
2 & 1\\
3 & 2\end{array}\right)\left(\begin{array}{c}
q\\
p\end{array}\right)\enskip\mathrm{mod}\ 1\label{eq:catmap}\end{equation}
 The Lyapunov exponent for this map is $\lambda=\ln(2+\sqrt{3})\approx1.317.$
The expanding and contracting eigenspaces through the origin are given
by $\xi_{u}=(-\sqrt{3},1)$ and $\xi_{s}=(\sqrt{3},1).$ The irrational
slope of the two directions implies that stable and unstable linear
manifolds are densely distributed over the torus.

The map is quantized by means of its generating function \cite{HannayBerry,QuantumMaps},
giving a unitary propagator $\hat{M}$ whose matrix elements in position
representation are\begin{equation}
\hat{M}_{j,k}=\frac{1}{\sqrt{N}}\exp\left[\frac{2\pi\mathrm{i}}{N}\left(j^{2}-jk+k^{2}\right)\right]\label{eq:quantumcat}\end{equation}

\subsection{Noise model}

\begin{figure*}[t]
\noindent \begin{centering}
\includegraphics[scale=0.85]{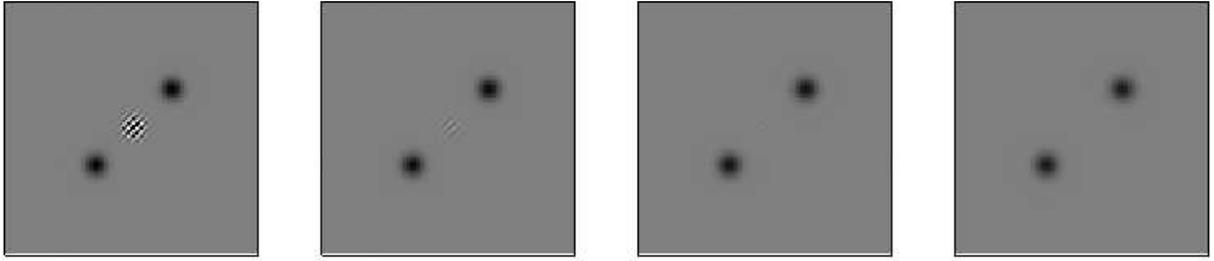}
\par\end{centering}

\raggedright{}\caption{\label{fig:figure1}Action of the noise model in phase space. 
The first panel (left) shows the discrete Wigner function of a superposition of two coherent
states centered at $(0.35,0.35)$ and $(0.65,0.65)$. The value of
the Wigner function is shown using a gray scale from white (minimum,
negative) to black (maximum, positive) (same scale for all panels). 
As time goes on (from left to right) the noise superoperator $\boldsymbol{D}_{\varepsilon}$
acting on the state washes out the interference fringes. We have taken $\varepsilon=0.05$ and $N=100$.}

\end{figure*}

\begin{center}
\begin{figure*}[t]
\noindent \begin{centering}
\includegraphics[scale=0.85]{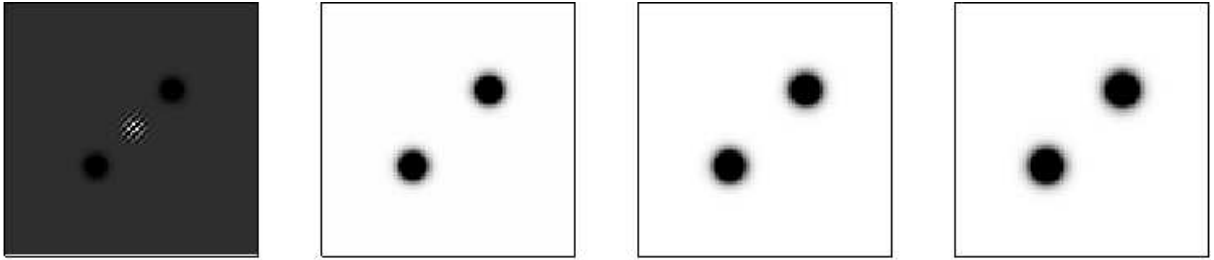}
\par\end{centering}

\raggedright{}\caption{\label{fig:figure2}Action of the noise superoperator on the same
initial state as in Fig. \ref{fig:figure1}, but with stronger coupling $(\varepsilon=0.1)$.
In this case the interference fringes disappear already in the first
application of the noise superoperator. The gray scale chosen in this case enhances the visualization
of the Wigner function spreading in phase space.}

\end{figure*}

\par\end{center}

We define the superoperator $\boldsymbol{D}_{\varepsilon}$ by means
of translation operators on the torus \cite{GarciaMata} 
\begin{equation}
\text{\ensuremath{\boldsymbol{D}}}_{\varepsilon}(\hat{\rho}_{t})=\sum_{j,k=0}^{N-1}c_{\varepsilon}(j,k)\ \hat{T}_{j,k}\ \hat{\rho}_{t}\ \hat{T}_{j,k}^{\dagger}\label{eq:diffusivenoise}
\end{equation}. These translation operators (our Kraus operators) are defined by Eq. (\ref{eq:translation_op}).
To preserve $\mathrm{Tr}(\hat{\rho}_{t})$ we assume $\sum_{j,k=0}^{N-1}c_{\varepsilon}(j,k)=1$
. This way, the coefficient $c_{\varepsilon}(j,k)$ represents the
probability of a translation being applied on the system in the direction
$(j,k)$. Defining this function as a periodized Gaussian (to satisfy
the boundary conditions on the torus) 

\begin{equation}
c_{\varepsilon}(j,k)\propto\sum_{\mu,\nu=-\infty}^{\infty}\exp\left[-\frac{(j-\mu N)^{2}+(k-\nu N)^{2}}{2\left(\frac{\varepsilon N}{2\pi}\right)^{2}}\right]\label{eq:coefficients}\end{equation}
we obtain a noise superoperator which has the effect of diffusing the
state on a region of radius $\approx\varepsilon$ in phase space.
The consequence of this incoherent superposition of translations is
decoherence, which can be visualized as the suppression of the small
scale interference fringes in the Wigner representation. After a short
time, the Wigner function becomes positive and the state appears {}``smeared
out'' in phase space. As previously mentioned, this behavior can be seen in Figs. \ref{fig:figure1}
and \ref{fig:figure2}, which show the effect of noise over the discrete
Wigner function \cite{Miquel,TracyPhD,OzorioRep} of a superposition
of coherent states. The parameter $\varepsilon$ can be interpreted
as a measure of the coupling between the system and the environment.

\section{Results}

In order to quantify the stability of the states of interest against
decoherence, we have calculated the purity \begin{equation}
P(t)=\mathrm{Tr}(\rho_{t}^{2})\label{eq:trace}\end{equation}
and the fidelity or autocorrelation function \begin{equation}
F(t)=\sqrt{\bra{\psi}\rho_{t}\ket{\psi}}\label{eq:fidelity}\end{equation}
as functions of time. In \eqref{eq:fidelity}, $\ket{\psi}$ represents
the pure initial state, hence $\rho_{0}=\ket{\psi}\bra{\psi}$. 

Purity is a measure of the correlation degree between the system and
the environment, and its evolution in time indicates how fast the
system loses coherence. Fidelity can be interpreted as the distance
between the evolved state and the initial state. Its development in
time allows us to measure the velocity with which the evolved state
{}``moves away\textquotedblright{} from the initial one under the
action of the noisy dynamics. Complementarily, this difference between
the initial and evolved states can be observed in terms of a density
operator representation in phase space, like Husimi or Wigner distribution
functions \cite{Schleich,OzorioRep}.

We have studied the evolution of these quantities from initial states
defined as scar functions, periodic orbit modes and eigenstates of
the unitary quantum map. For each system, we have built the scar functions
and the periodic orbit modes on different short periodic orbits of
the corresponding classical map (without noise) and then compared
their evolution to those of the map eigenstates, in particular with
those localized on or nearer the same orbits. We have also studied the
effect of varying the coupling with the environment by means of taking 
different values for the parameter $\varepsilon$.

In the following Figures, we show some typical results of our numerical
calculations. They are illustrative of an exhaustive exploration 
of the eigenstates of our systems. Fig. \ref{fig:baker_figs} shows the behavior of purity
and fidelity for the case of the noisy baker map model. The states
are localized over a period 2 orbit with symbolic code $01$, as
can be seen in the Husimi representations of the upper panel. The
overlap between the scar function and the map eigenstate is $0.828$,
so the states are quite similar in terms of this distance measure.
However, there is a visibly different behavior in terms of purity
and fidelity, favoring the semiclassically constructed states in general,
and the scar functions in particular, which lose purity and fidelity
at a slower pace. 

For the noisy cat map we have a similar behavior, but the differences
between states in terms of purity and fidelity evolution are less
pronounced, as can be seen in Fig. \ref{fig:cat_figs}. In this case,
the scarring of map eigenstates by periodic orbits is not so strong 
as in the baker map case. However, scar functions are more robust than
the other states, as in the previous model. 

These results show that this behavior is independent of the kind of 
noise considered, when compared to what has been shown in \cite{Raviola}. There, 
a different open system was studied (a dissipative baker map), in which the noise
was non-generic because it acts along a preferential direction in
phase space, corresponding to the stable manifolds of the classical map. 
Also, there is no dependence on the kind of map. In fact, 
in a more generic system in terms of scarring like the cat map, the 
same has been found. 

In order to propose an explanation for the behavior of scar functions, 
in Figs \ref{fig:bakerevol} and \ref{fig:catevol} we show the Wigner distributions corresponding to 
the first $4$ steps of the evolution of the states used in Figs. \ref{fig:baker_figs} and 
\ref{fig:cat_figs}, respectively.  In the upper panels we see the eigenstates of the maps, which have a 
complicated background structure along with localization on the corresponding orbit. 
The scar functions and POMs have a much simpler shape, with details 
that in general live longer than those present in the eigenstates. In fact, the eigenstates 
seem to converge to the corresponding scar functions. This is the 
underlying mechanism that produces a faster loss of fidelity and purity in the eigenstates with 
respect to the classically motivated quantum distributions. 
Finally, a brief discussion about the similar behavior of the decay of 
the purity and the fidelity. The first quantity measures the rate of coherence loss, that can be 
seen very clearly through the Wigner distributions. The second one measures the correlation 
between the initial state and the evolved ones. The interaction with the environment essentially destroys 
the interference fringes whilst the dynamics distorts the initial distributions. But all three 
initial states are very localized on a periodic orbit and its manifolds. Then the main initial contribution 
to the loss of both, purity and fidelity comes from the destruction of coherences.

\begin{center}
\begin{figure*}
\begin{centering}
\includegraphics[scale=0.6]{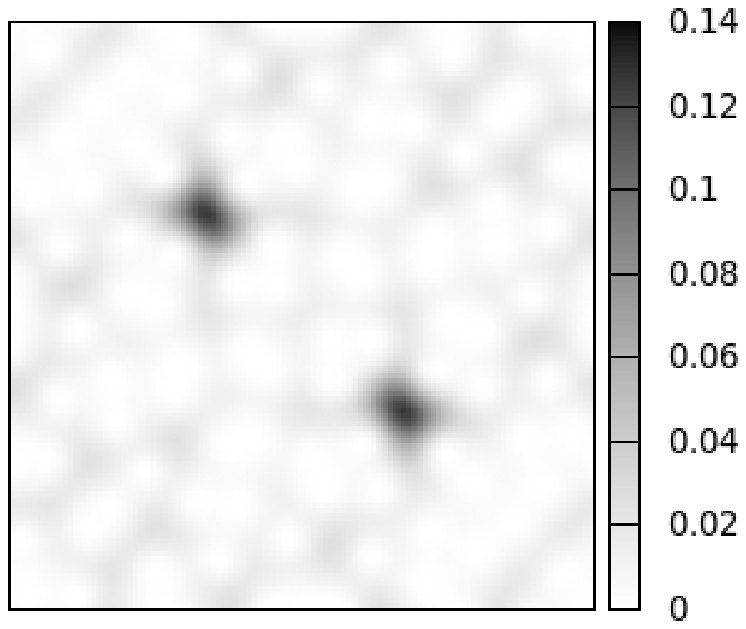}\includegraphics[scale=0.6]{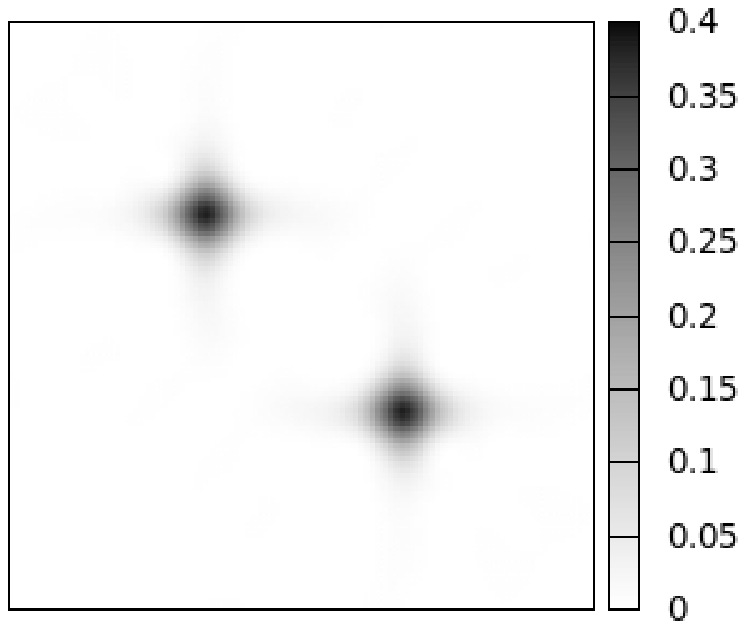}
\par\end{centering}

\includegraphics[scale=0.4, angle=-90]{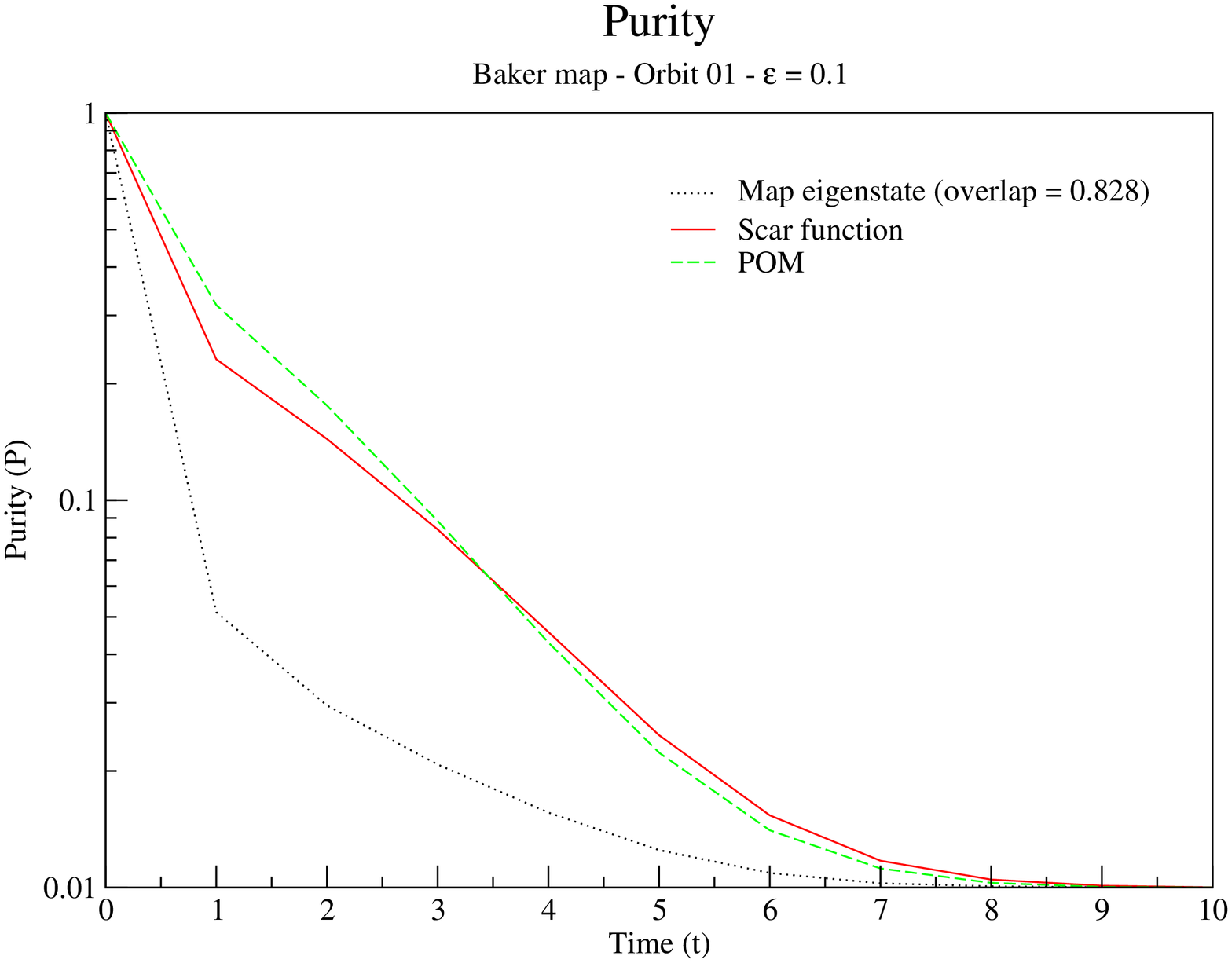}

\includegraphics[scale=0.4, angle=-90]{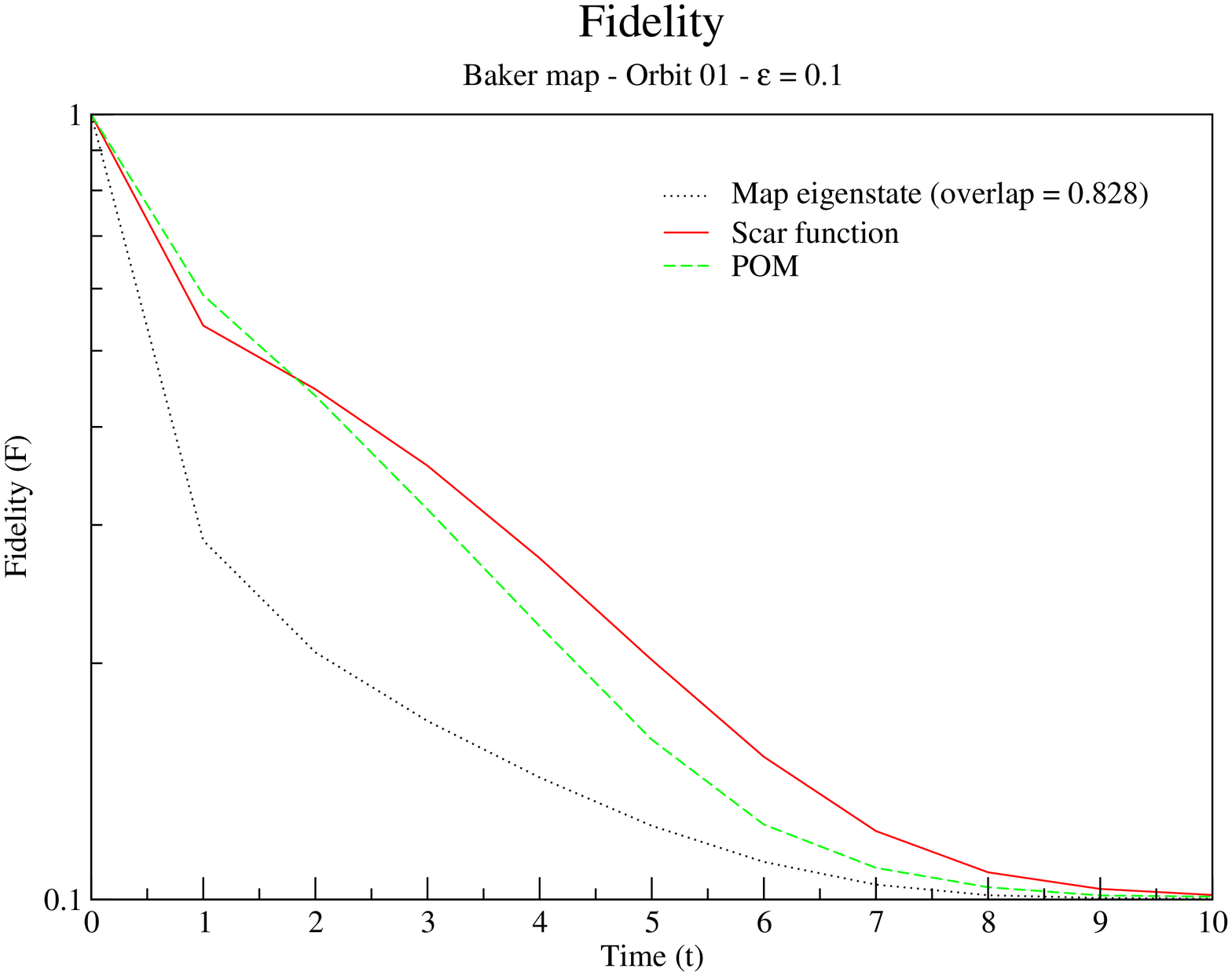}

\caption{\label{fig:baker_figs}Purity and fidelity behavior for the noisy
baker map. In the upper panels we show the Husimi representation
of two initial states, localized near the period-2 orbit 01. Black 
corresponds to maximum probability, and white to minimum. Upper left panel: map eigenstate.
Upper right panel: scar function. Middle panel: purity evolution (logarithmic
scale). Lower panel: fidelity evolution (logarithmic scale). In the last two panels, black dotted
lines correspond to the map eigenstate, green dashed lines to the POM, and red solid lines to the 
scar function. We have taken $\varepsilon=0.1,\ N=100$.}

\end{figure*}

\par\end{center}

\begin{center}
\begin{figure*}
\begin{centering}
\includegraphics[scale=0.45]{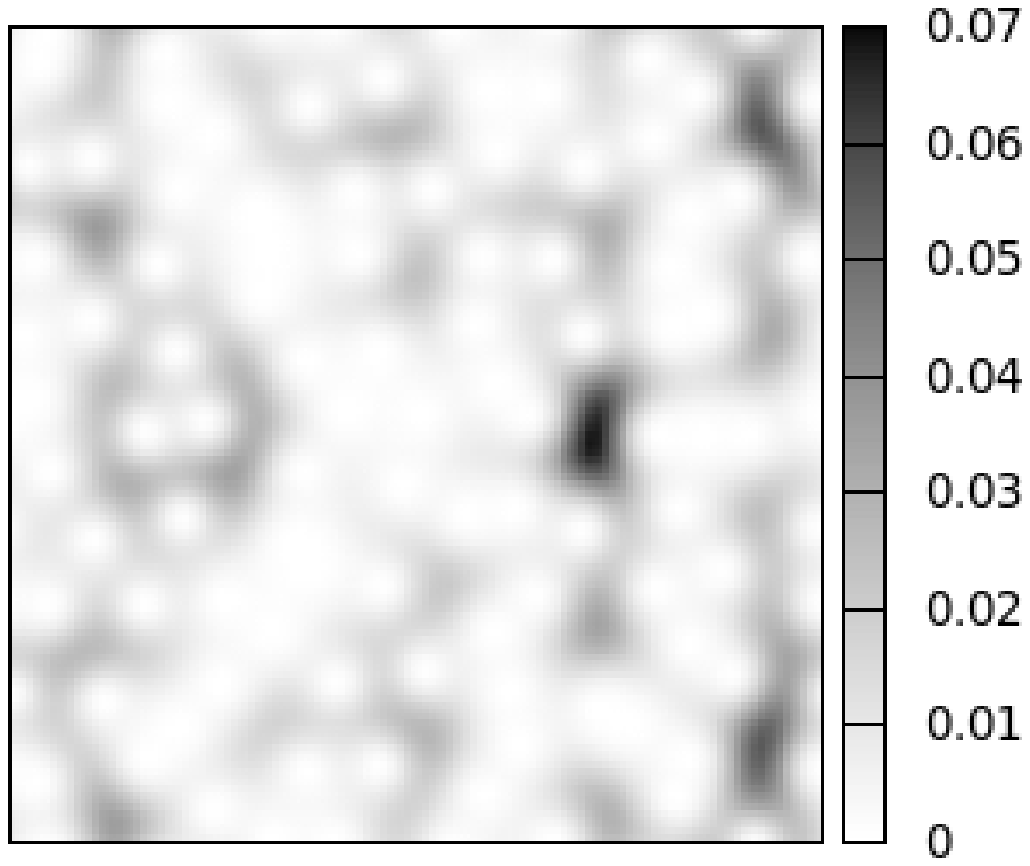}\includegraphics[scale=0.45]{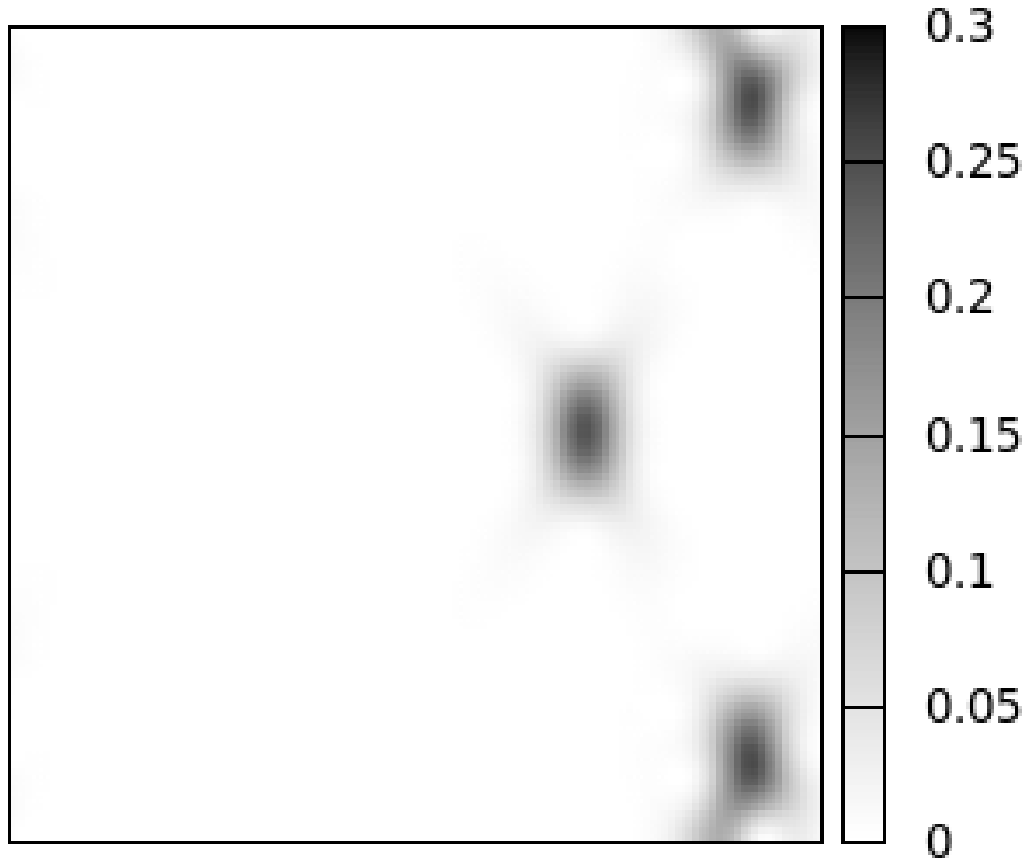}
\par\end{centering}

\includegraphics[scale=0.4, angle=-90]{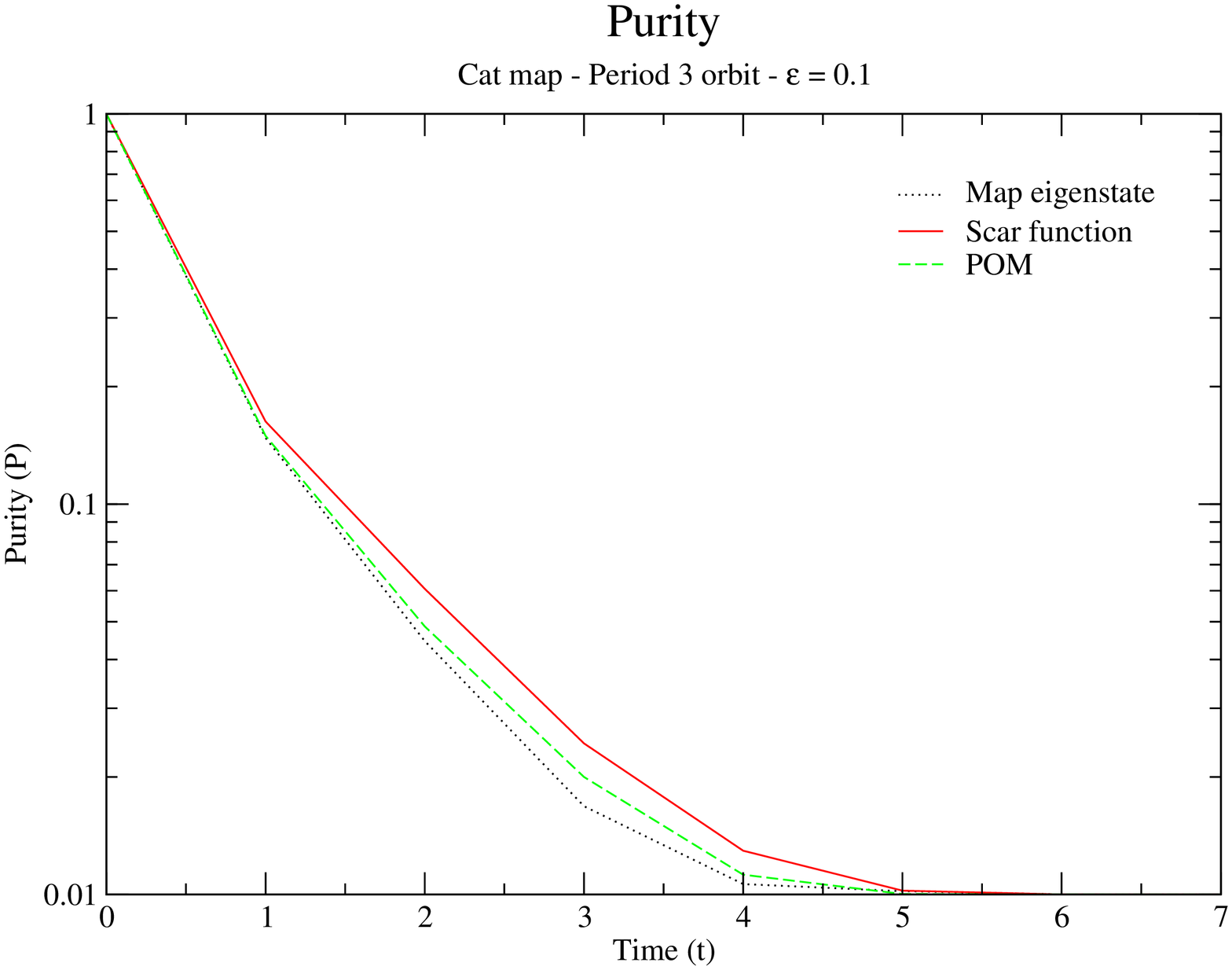}

\includegraphics[scale=0.4, angle=-90]{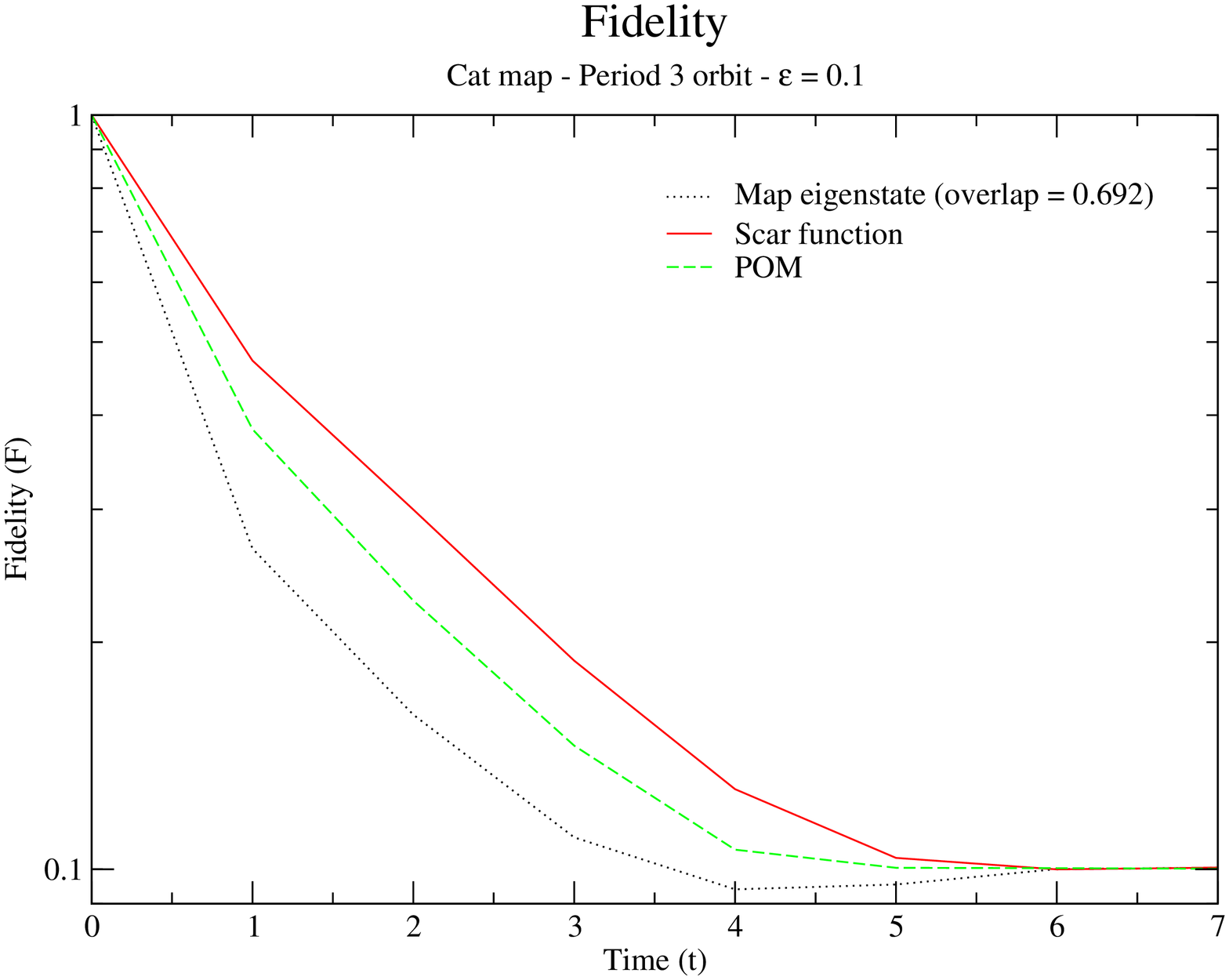}

\caption{\label{fig:cat_figs}Purity and fidelity behavior for the noisy cat
map. In the upper panels we show the Husimi representation of two initial
states, localized near a period-3 orbit. Colors, patterns, scales and parameters
as in Fig. \ref{fig:baker_figs}.}

\end{figure*}

\par\end{center}

\begin{center}
\begin{figure*}
\begin{centering}
\includegraphics[scale=0.8]{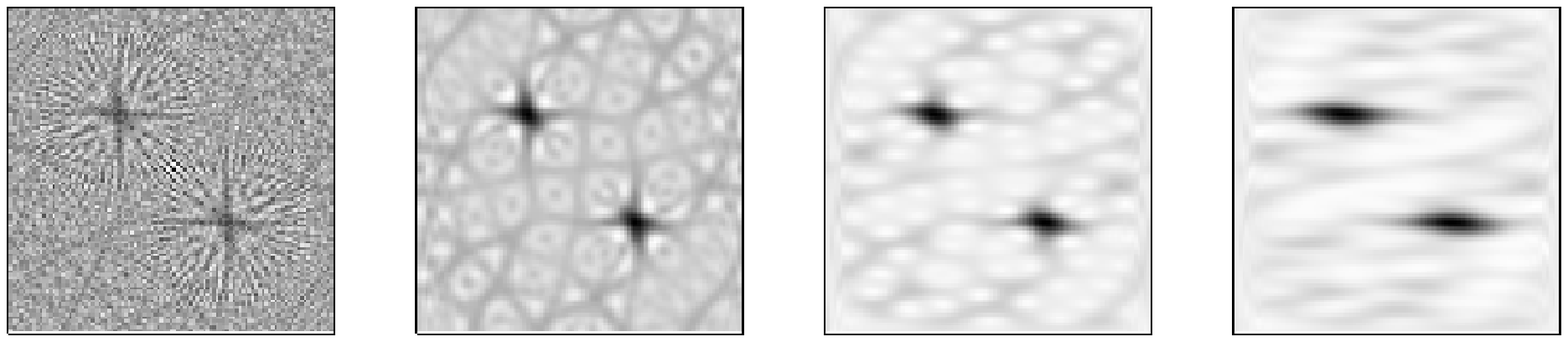}
\includegraphics[scale=0.8]{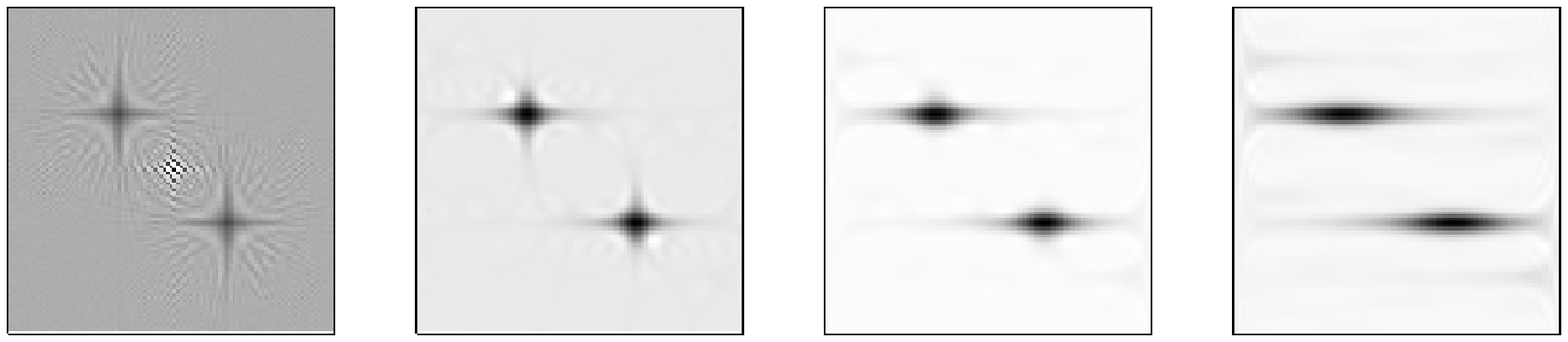}
\includegraphics[scale=0.8]{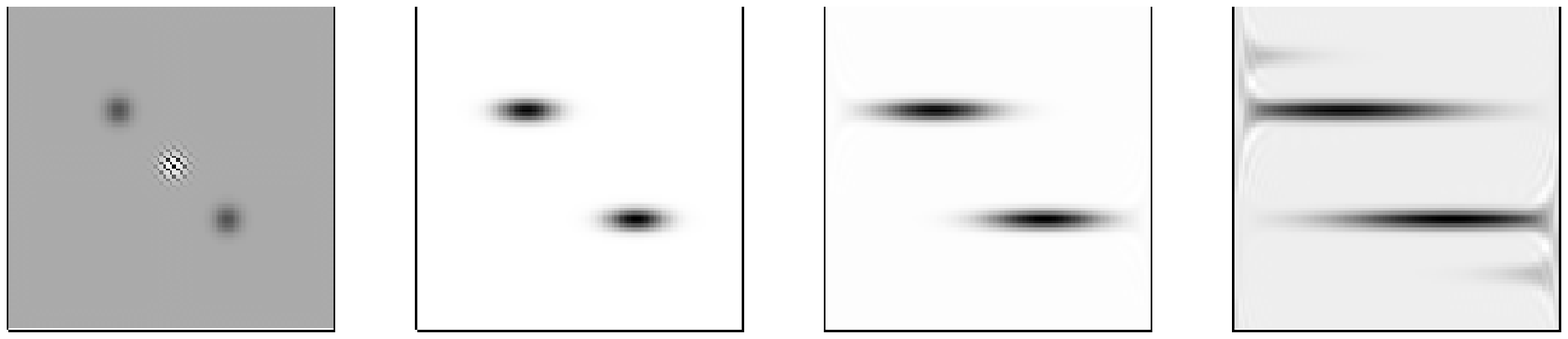}

\par\end{centering}

\caption{\label{fig:bakerevol} Wigner distributions corresponding to the first $4$ steps of evolution 
of a baker map eigenstate (upper panels), the corresponding scar function (middle panels), 
and POM (lower panels). We have taken the same parameters as those chosen 
for Fig. \ref{fig:baker_figs}.}

\end{figure*}

\par\end{center}

\begin{center}
\begin{figure*}
\begin{centering}
\includegraphics[scale=0.8]{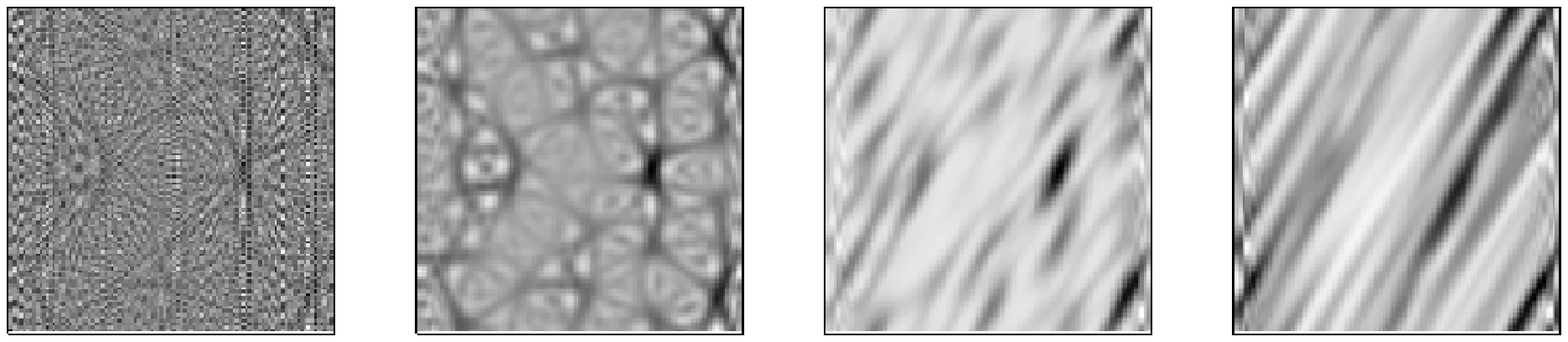}
\includegraphics[scale=0.8]{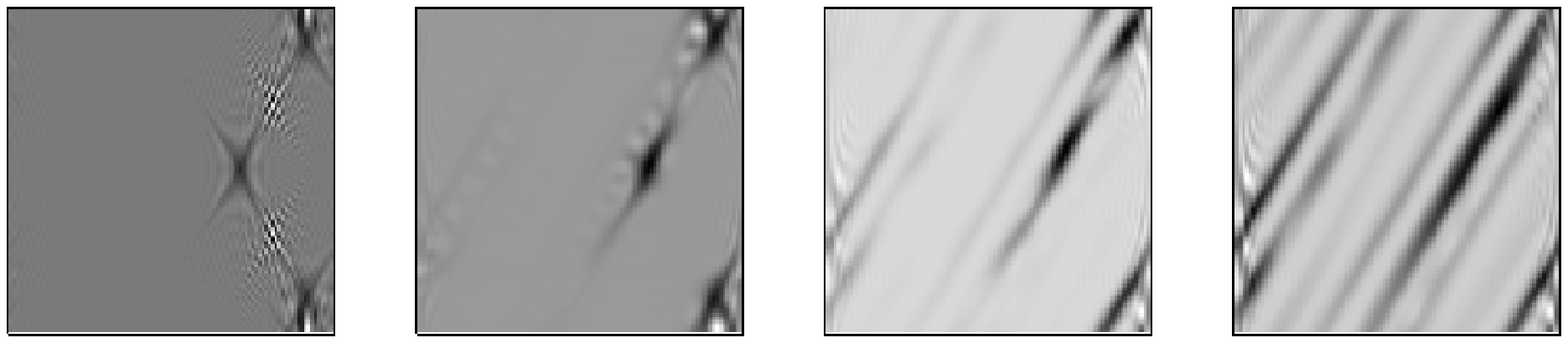}
\includegraphics[scale=0.8]{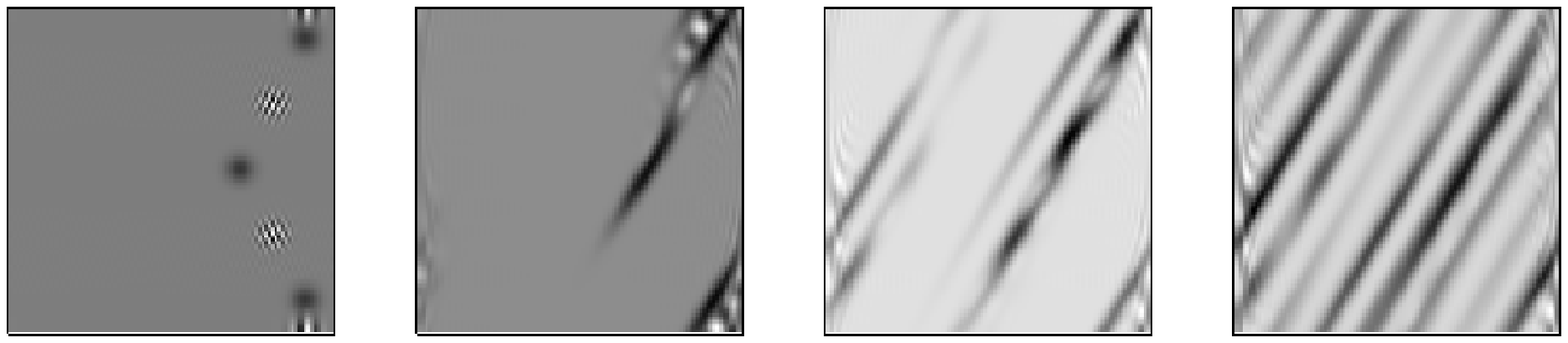}

\par\end{centering}

\caption{\label{fig:catevol} Wigner distributions corresponding to the first $4$ steps of evolution 
of a cat map eigenstate (upper panels), the corresponding scar function (middle panels), 
and POM (lower panels). We have taken the same parameters as those chosen 
for Fig. \ref{fig:cat_figs}.}

\end{figure*}

\par\end{center}

\section{Summary and conclusions}

We have studied two complementary decoherence measures purity
and fidelity for a generic diffusive noise in two different chaotic maps.

For both quantities, we have found classical structures in quantum mechanics
that are specially stable when subjected to environmental perturbations.
They are the scar functions, which are associated to periodic orbits
and the stable and unstable manifolds in their vicinity.

We have seen that quantum states constructed on classical invariants, 
periodic orbits and their stable and unstable manifolds,
are more stable against an external noise than the eigenstates of the closed 
quantum system. This turns them into the most robust significant quantum 
distributions in generic dissipative maps. We conjecture that 
the scar functions will be the most robust structures also in general chaotic 
systems. 

This result has already been announced for a particular dissipative noise
in the Baker map. But in that case the noise was along the stable direction 
of the hyperbolic structure of the original map. Here, we have confirmed 
the same result for a general diffusive model for two different maps.

We can then say that the external noise destroys the stability of the
quantum invariants faster than the stability of the
classical ones. This is a consequence of the effect of the noise as 
it quickly destroys quantum interferences whilst only spreads the classical
structures in phase space. Despite we have only shown results for 
$\varepsilon=0.1$, we have verified that they represent the generic behavior for 
a wide range of couplings. For strong couplings ($\varepsilon>1$) the noise 
destroys the distributions very fast, and for weak ones ($\varepsilon<0.001$) 
the system behaves in a similar way to the closed one.

Then, we have shown that for generic maps scar functions represent the
stable classical skeleton of the map eigenstates against environmental
perturbations.

We are currently developing the theory to quantitatively explain this behavior. 

\begin{acknowledgments}
Support from CONICET is gratefully acknowledged.\end{acknowledgments}

\end{document}